# DRAFT

# Super Earth Explorer:

# A Coronagraphic Off-Axis Space Telescope

Schneider J.[a], Boccaletti A.[b], Mawet D.[e] Baudoz P.[b], Beuzit J.-L.[c], Doyon R.[d] Marley M.[e], Stam D [g]., Tinetti G[h]. Traub W. [f], Trauger J.[f], Aylward A.[h], Cho J. Y-K. [i], Keller C.-U. [J], Udry S.[k], and the SEE-COAST Team [1]

*[a] LUTH, Paris Observatory; [b] LESIA, Paris Observatory; [c] LAOG; [d] U. Montreal, [e] NASA/Ames, [f] Jet Propulsion Laboratory, California Institute of Technology; [g] SRON; [h]UCL, ; [i] Queen Mary, U. London, [J] U. Utrecht,, [k] Geneva Observatory*

jean.schneider.obspm.fr

1 Full list at the end of the paper

Abstract:

*The Super-Earth Explorer is an Off-Axis Space Telescope (SEE-COAST) designed for high contrast imaging. Its scientific objective is to make the physico-chemical characterization of exoplanets possibly down to 2 Earth radii . For that purpose it will analyze the spectral and polarimetric properties of the parent starlight reflected by the planets, in the wavelength range 400-1250 nm*

# 1. Scientific Objectives

The currently known planetary systems from indirect detection show a huge diversity in the orbits (semi-major axis and eccentricity), masses, and, in a few cases, radii of their planets. If we add to this the observation that the planets in our Solar System differ strongly from each other, there is no doubt that more information on the physical characteristics of exoplanets will reveal not only new and unexpected scenarios but also answers and a better understanding. SEE-COAST will make a first and highly anticipated inventory of these physical characteristics.

## 1.1 The need for direct detection of exoplanets

Today more than 300 exoplanets have been found by radial velocity , transits and microlensing (Schneider 2008) and more will be detected by astrometry (Sozzetti et al. 2007). These indirect detection methods provide little to no information about the physical characteristics of a planet, such as the composition and structure of its atmosphere and the properties of its surface (if there is any). In order to characterize the physical properties of exoplanets, one needs to directly detect planetary radiation. This is precisely what SEE-COAST is designed for.

Planetary emission consists of visible to near-infrared reflected starlight and of intrinsic infrared thermal radiation. Both the reflected flux and the thermal flux depend on the size of the planet, on the distance between the observer and the planet, and on the planet's phase angle (i.e. the angle between the star and the planet as seen from the observer). They also depend on the composition and structure of a planet's atmosphere, the properties of the surface (if applicable), the wavelength, and e.g. the presence and physical characteristics of planetary rings and/or large companions. This is why planetary emission can be used to characterize planets. Characterization of exoplanets by space missions in the thermal infrared regime is for instance discussed in Cockel et al (2008). Here we describe the mission SEE-COAST devoted to the reflected light in the visible.

## 1.2The diversity of planets

### *1.2.1Giant planets*

Nature forms planets very different than those of in our Solar system. We already know of gas giants possessing cores of 70 $M_{Earth}$, as typified by HD 149026b (Sato et al., 2005; Fortney et al., 2006).

Given that migration appears to be common in exoplanetary systems, there is no reason to think that ' ice-rich giant' planets, like Uranus and Neptune, should be relegated to the cold outer fringes of planetary systems. Kuchner (2003) and Léger et al. (2004) have postulated that volatile-rich planets with masses from ~1 to ~15 $M_{Earth}$ may well reside at a great variety of orbital distances. Further work on core-accretion planet formation models (Ida & Lin, 2004; Benz 2006) suggests that massive ice-rich planets may be extremely common in extrasolar systems, likely outnumbering the gas giants.

Fortney et al. (2007) have calculated the internal structure and radius of iron-, rock-, and ice-rich planets as a function of mass and composition. Ice-rich planets have a radius of 3 to 4 $R_{Earth}$ and, given a warm orbit (a few AU), their atmospheres will sport bright water clouds. Such planets could easily be almost as bright in reflected light as a conventional Jupiter at the same orbital radius.

### *1.2.2 Super-Earths*

A 'super-Earth' is a planet larger than the Earth that is not dominated by an atmosphere (like gas giants are). Its radius ranges from about 1.5 to 2.5 $R_{Earth}$. Several super-Earth planet families can be considered, including ones unknown in the Solar system: large 'Mercury-like' planets, with an iron core and a thin layer of silicates, 'Venus/Earth/Mars-like' planets, with an iron core and a thick silicate mantle, 'water'- or 'ocean'-planets made of iron, silicates, and water (similar to icy moons of Jupiter and Saturn), 'carbon'-planets, and combinations of the above (Sotin, Grasset & Mocquet, 2007; Valencia et al. 2007; Fortney et al., 2006; Léger et al., 2004; Kuchner, 2003).

In the quest for life in the universe, we have to be open-minded, and should not limit ourselves to the, conservative, search for an 'Earth's sister'. Indeed, planets significantly smaller than the Earth have probably lost the biologically interesting material needed for the evolution of life. Planets larger than the Earth, that have the operational advantage of being easier to detect, appear to be more promising because they can better retain water and an atmosphere. The atmospheric dynamics should not be dramatically affected by the change of planetary scale. ...However, one parameter that plays a significant role in the dynamics is the planet's rotation period.

## 1.3 How frequent are super-Earths?

The current exoplanet detection techniques are mostly sensitive to planets larger than super-Earths. The lowest achieved limits are about 5 Earth masses (Rivera et al., 2005; Beaulieu et al., 2006; Udry et al., 2007) but both observational evidence and theoretical simulations indicate that small planets are more numerous than giant ones: the observed mass histogram rises towards small masses.

Results of existing radial velocity surveys show that at least 30% of G-K stars have at least one super-Earth with a mass less than $25M_{Earth}$ and an orbital period less than 50 days (Mayor et al. 2008).

Recent planet formation models predict that super-Earth planets indeed exist at various distances from their parent stars (Mordasini et al. 2007). These models predict a substantial migration extent for the planets, which therefore are likely to start their formation beyond the ice-line. Hence, they contain a large amount of water in the form of a standard or a super-critical liquid. In the second type of models, super-Earths form in a way similar to terrestrial planets, from gravitational interactions between many different embryos. Radial mixing enables water to be present in planets at small distances from the central star, and planets with giant oceans are to be expected.

Terrestrial exoplanets have already been detected. Three small terrestrial planets have already been detected around a non-main sequence star (PSR 1257+12).

Summarizing, all types of planets may exist at all distances in other planetary systems, and about 50% % of stars may have Earths or super-Earths at a distance of about 1 AU.

## 1.4 What SEE-COAST will observe

With SEE-COAST, we propose to measure the starlight that is reflected by exoplanets, from the far blue (0.4 µm) to the near-infrared (1.25 µm) with a spectral resolution of 40 to 80. Not only will we measure the spectral distribution of the *flux* of this reflected light but also that of its degree and direction of linear *polarization*.

Spectroscopy provides very rich information, such as the species present in the atmosphere, planet's cloud coverage and, if applicable, its surface properties. Polarimetry is a strong tool for the detection and characterization of exoplanets. The reflected starlight is generally polarized because of e.g. scattering within the planetary atmosphere, and can thus be distinguished from the unpolarized direct starlight (Kemp et al. 1987). In addition, the state of polarization of a planet depends strongly on its physical characteristics (Seager et al. 2000, Stam et al. 2004), just like the total reflected flux.

SEE-COAST will focus on exoplanets that can be spatially resolved from their star. The planet's radiation will be observed, while most of the starlight is blocked by the coronagraph. Its purpose is not, primarily, to search for new planets but to observe planets previously known from radial velocity and astrometric surveys. The main science drivers of the SEE-COAST mission are:

**Gaseous planets** with orbital distances from 1 to 5 AU, around stars up to 10 pc away. We anticipate a huge variation in atmospheric compositions and "accessories" such as large companions and planetary rings.

**"Super-Earths"** (i.e., terrestrial planets with radii from 1.5 to 2.5 Earth radii) in or close to the habitable zone of nearby stars. If these planets have oceans and a relatively transparent atmosphere, SEE-COAST can make the first direct detection of liquid water on an exoplanet.

**Planetary systems as a whole**; SEE-COAST can provide information about the parent star, detect and characterize multiple planets in a system, an exo-zodiacal disk, and reservoirs of exo-comets, like the Kuiper belt in the Solar system.

With SEE-COAST we can access an unprecedented diversity of exoplanets and exoplanet characteristics that will help us understand and evaluate not only exoplanetary systems but also our own Solar system.

## 1.5 What SEE-COAST will characterize

### *1.5.1Atmospheric species*

Table 1 lists the atmospheric gas molecules with absorption bands in the visible and near-infrared that SEE-COAST will be able to detect and for which it can derive mixing ratios.

Water, oxygen, methane, carbon dioxide, and ozone give the key signatures and determine the primary requirements for SEE-COAST's baseline performance.

Water ($H_2O$) is made from the two most abundant chemically reactive elements in the universe, and it is the necessary ingredient for the types of life found on Earth. For planets at orbital distance 1 AU, $H_2O$ is expected to be one of the most abundant atmospheric components. Liquid water has played an intimate, if not fully understood, role in the origin and development of life on Earth. Water contributes to the dynamic properties of a terrestrial planet, permitting convection within the planetary crust that might be essential to supporting Earth-like life by creating local chemical disequilibria that provide energy. Water absorbs electro-magnetic radiation over a broad wavelength range, covering part of the visible and most of the near-IR, and has a very distinctive spectral signature (see Table 1). Abundant water may be likely on some planets since it was detected on HD 189733 b (Tinetti et al. 2007, Swain et al 2008).

In addition to water, the search for carbon-dioxide ($CO_2$) is of special interest (see Table 1). Its presence would indicate (1) that carbon is available for the biosphere, (2) a (natural) greenhouse effect, and (3) a possible regulation by the hydro- and geosphere.

The greatest surprise in the composition of the planets in our Solar System, is the large amount of oxygen ($O_2$) in the terrestrial atmosphere. This molecule is so reactive chemically that it must be continuously produced at enormous rates to persist. Thus the Earth's atmosphere can only be the result of a large input from the biosphere. The challenge of remotely detecting life on a planet that has not developed a biogenic source of oxygen is fraught with unknowns. $O_2$ shows a spectral signature only in the VIS wavelength range (see Table 1 for the location of the strongest band, the so-called oxygen *A*-band).

**Table 1:** Absorption bands of the most important atmospheric molecules that SEE-COAST can detect and identify in the visible/near-infrared.

| **Molecule** | **Wavelength (micron)** |
|---|---|
| $O_2$ | 0.76 |
| $H_2O$ | 0.51, 0.57, 0.61, 0.65, 0.72, 0.82, 0.94, 1.13 |
| $CH_4$ | 0.48, 0.54, 0.57. 0.6, 0.67, 0.7, 0.79, 0.84, 0.86, 0.73, 0.89 |
| $CO_2$ | 1.25 |
| $NH_3$ | 0.55, 0.65, 0.93 |
| $O_3$ | 0.45-0.75 (the Chappuis band) |

### *1.5.2 The planetary albedo*

The planet's albedo will be derived from the reflected flux by assuming a planet radius. For massive (down to Saturn-mass) planets, all models agree that the radius is very close to 1 Jupiter radius. For lighter planets, we make use, to first order, of the mass-radius relations given by Grasset et al. (2007); Valencia and Sasselov (2006); Fortney et al., (2007). If a planet is known from radial velocity observations, its mass can be derived from these observations when combined with the observed orbital inclination angle. Depending on a planet's composition, its radius can vary by a factor of up 20% for a given mass. This translates into an uncertainty of 40% in the albedo. The value of albedo can then be refined using flux and polarization spectra, the time variability, and theoretical models.

### *1.5.3 Surface features and planet rotation from albedo time variability*

SEE-COAST will measure the time-variability of flux and polarization spectra. This variability could arise from daily rotation, with different parts of the planet rotating into view, from the orbital movement of the planet, which will generally change the phase angle, or from delayed impact of stellar activity. There might also be time-dependent changes on the planet itself, such as due to weather and seasons, leading for instyance to diurnal flux changes (Ford et al. 2001) and diurnal polarization changes (Stam et al., 2004). Similar events could occur on super-Earths. Time variability in spectra can provide information not only about planetary rotation rates, but also about the distribution of surface types, aerosol and clouds. (Palle et al . 2008 and Williams and Gaidos 2008). To capture a planet's time variability, SEE-COAST will visit it several times, with a total integration time up to several weeks.

### *1.5.4 From albedo to internal dynamics*

Relationships between atmospheres and internal dynamics of planets are not clearly understood for Solar System bodies and have not been investigated for super-Earths. Atmospheric composition depends on (1) the intensity of volcanic and tectonic activity, (2) the oxidization degree of the silicate mantle which will constrain both nature and quantity of gases that are expelled, and (3) the age of the planet. SEE-COAST will provide valuable constraints on these questions. Since it will discover new types of planets, while yielding the main composition of their atmospheres, it will add new data to what we know from solar system bodies. Such data are fundamental for a more thorough understanding of the relationships between a planet's internal activity and atmospheric composition.

## 1.6 Requirements

The science requirements as derived from SEE-COAST's science objectives are the following :

- The instrument should be sensitive to planets sizes down to 2.5 Earth radii at distances of 2.5 pc and down to 1 Earth radius for very close stars. This sensitivity amounts to e.g. 3000 photons/day/spectral resolution element (R=40) for a 2.5 Earth-radii planet at a distance of 2.5 pc.
- We are interested in minimum star-planet separations of 0.8 AU. This requires an angular resolution of **70 mas.**
- In order to visualize planetary systems with planets up to 5 AU at distances down to 3 pc, a field of view (FOV) of 3'' in diamater is needed.
- To measure both Rayleigh scattered light and an absorption band of $CO_2$, a wavelength range from 0.4 to 1.2 micron is needed. Our goal is to go down to 0.3 micron.
- The identification of chemical species, and in particular the signatures of biologically interesting molecules, requires a spectral resolution of 40 in the VIS, 80 in the NIR.
- In order to use polarization measurements for planet characterization, an accuracy of 10% is needed in the degree of polarization for a planet magnitude of 25.
- To determine the orbit and capture seasonal variations on a planet, it is necessary to observe the planet 5 to 10 times along its orbit. It should therefore be visible about 6 months per year.

These requirements are summarized in Table 2.

**Table 2:** SEE-COAST's Scientific objectives.

| Science constraint | Value |
|---|---|
| Minimum planet radius at 3 pc | 2.5 $R_{Earth}$ |
| Star-planet distance | 0.8– 5 AU |
| FOV to detect planet at 5 AU down to 3 pc | 3 arcsec (diameter) |
| Spectral range to detect Rayleigh scattering and CO2 | 0.4 – 1.25 micron |
| Spectral resolution to detect $CH_4$, NH3, H2O , O2 in the visible | R = 40 |
| Spectral resolution to detect $CO_2$ in the NIR | R = 80 |
| Minimum detectable depth of spectral features | 30% |
| Minimum detectable polarization | 30% |
| Time span of target visibility | 6 month/yr |

The requirements for the telescope and the spacecraft, as derived from SEE-COAST's science objectives and instrument requirements, are the following .

- A sensitivity of 3000 photons/day/spectral resolution element (R=40), as well as the angular resolution of 70 mas, require a pupil aperture of at least 1.5 m.
- At the inner working angle of 2 lambda /D, the residual stellar speckle should be calibrated and subtracted at a level sufficiently low to extract the 1:10e9 planet signal.
- The telescope will be placed at the L2 Lagragian point. This choice results from several requirements. First the extreme sensitivity of speckle stability to temperature fluctuations requires a very high thermally stable environment. Second, SEE-COAST needs to monitor the most interesting planets for several weeks without interruption (other than downloading of data). Finally, L2 is well suited for a full sky coverage as exoplanets are distributed randomly in the sky. These characteristics are summarized in Table 3.

**Table 3:** Payload & spacecraft characteristicss.

| **Parameter** | **Value** |
|---|---|
| Entrance pupil diameter D | > 1.5 |
| Angular resolution | 70 mas @ 0.6 micron |
| Contrast (after speckle subtraction) @ 2 lambda/D | $<10^{-8}$ |
| Contrast (after speckle subtraction) @ 4 lambda/D | $<10^{-9}$ |
| Orbit for 6 months visibility | L2 Lagrangian point |

# 2.Problems of direct imaging

SEE-COAST is designed for directly detect exoplanets against the stellar background. This background has not the traditional structure as in classical imaging (mainly sky background) but is composed of the diffracted light of the on-axis source, i.e. the star. This Point Spread Function has a deterministic part (the diffraction of the telescope pupil) and a speckled halo mostly resulting from phase aberrations. Actually, the complex PSF structure is also affected by amplitude aberrations and stability of the instrument. Coronagraphy is a technique which aims to reduce the light diffracted by the telescope pupil while wavefront correction and speckle calibration are designed to address the second effect.

Speckle calibration refers to the use of the differentiation between the speckles (originating from the on-axis star) and the planets (off-axis objects). That can be, for instance, a difference in the spectral or polarimetric signals of speckles vs.

planets. The focal instrument will include the required elements to take advantage of these effects to improve performance.

A combination of several high contrasts imaging technique will be required for SEE-COAST to achieve the huge contrast between a star and its planet.

# 3.Payload

## 3.1Overview

SEE-COAST can achieve the scientific goals discussed in detail in section 1, using an off-axis, 1.5 meter telescope on an orbit around the L2 Earth-Sun Lagrange point, coupled to a coronagraph that feeds an imaging spectro-polarimeter. The payload has two main components (Fig. 1):

- An off-axis telescope with a 1.5-meter prime mirror and 3 star guiders.
- Two focal plane instruments:

1. An imaging spectro-polarimeter behind a coronagraph, with a FOV of 3’’x3’’. It has two channels: a visible (0.4 – 0.85 mm), and a near-infrared (0.85 – 1.25 micron).

2. A pick-up mode camera with a FOV of 30’’x30’’.

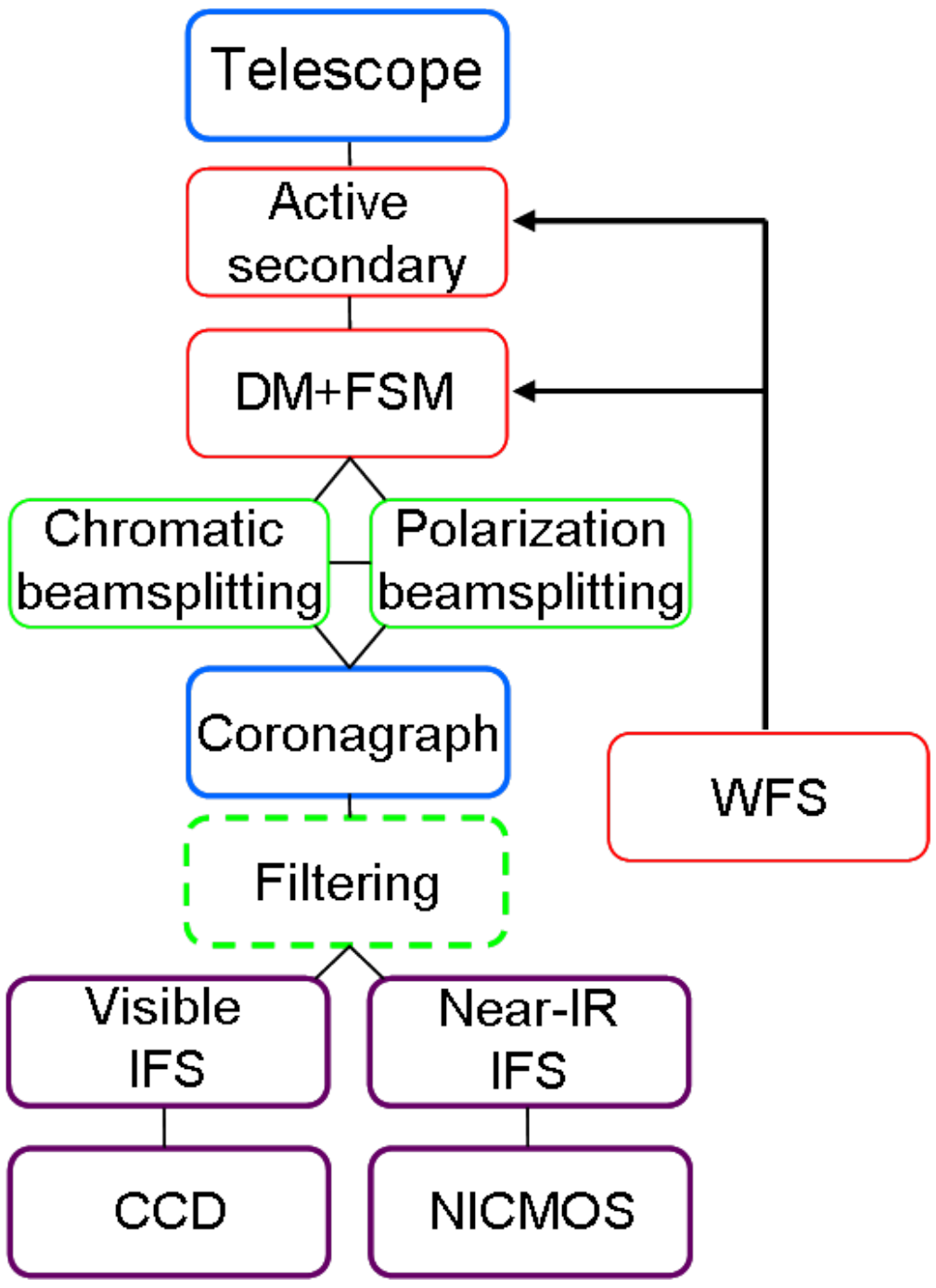

**Figure 1:** Block diagram of the main payload elements

Figure 1 shows the block diagram of the main payload elements. The off-axis telescope is actively corrected to provide a stabilized image of excellent quality to the main high-contrast spectro-polarimetric camera. The fine pointing will be done with a parabolic mirror and a tip-tilt mirror (Fast Steering Mirror, FSM). Finally, an off-axis parabola is used to obtain a collimated beam as input for the spectro-polarimeter. The beam is divided in 2 spectral channels. A modulator then selects the proper polarization direction well-in-sync with the detector. The coronagraph suppresses the coherent part of the stellar light with the help of a filtering stage composed of a classical spatial filter (a Lyot stop) and an optional polarization filter. The residual signal still hiding the exoplanet light is then sent to a lenslet-based integral field spectrograph with a relatively low spectral resolution (R=40-80).

The top-level requirements on the payload are the angular resolution (primary mirror diameter of 1.5-2 m), low wavefront bumpiness (WFE < $\lambda$/100 @ 633 nm, high contrast at small separations ($10^{-7}$-$10^{-8}$ at 1-2 $\lambda$/D before speckle nulling, subtraction and calibration), high stability (L2 orbit, duration on the order of 1 day) and the ability to have full spectral coverage in the visible and near-infrared simultaneously.

### 3.2 **The telescope**

The proposed SEE-COAST telescope concept is an unobstructed off-axis Cassegrain (Fig. 2). Stringent fabrication and control of optical surfaces and alignments is needed in order to preserve the quality of the image as it is sent to the science instruments (see Table 4)..

**Table 4:** Telescope design parameters.

| | |
|---|---|
| Primary mirror F-number | F/60 |
| Entrance aperture | 1.5 m |
| Image quality over 3" | lambda/100 rms @ 633 nm |
| Image quality over 30" | lambda/20 rms@ 633 nm |

Two options are identified for the telescope. The first one is to have a very high quality super-polish telescope (5 nm-rms in the mid-spatial frequency range 5-40cm/cycle) to reduce as much as possible the level of the speckled halo downstream the coronagraph.

Second option is to relax the optical quality of the primary mirror but to allow active wavefront correction at the mid-frequencies with a deformable mirror (DM). This has to be done at room temperature to allow the DM to work.

Behind the focus, a parabolic mirror and a tip-tilt mirror (FSM) are planned to be used as an active system to finely correct the telescope pointing defects. Finally, on off-axis parabola is used to obtain a collimated beam as input for the spectro-polarimeter.

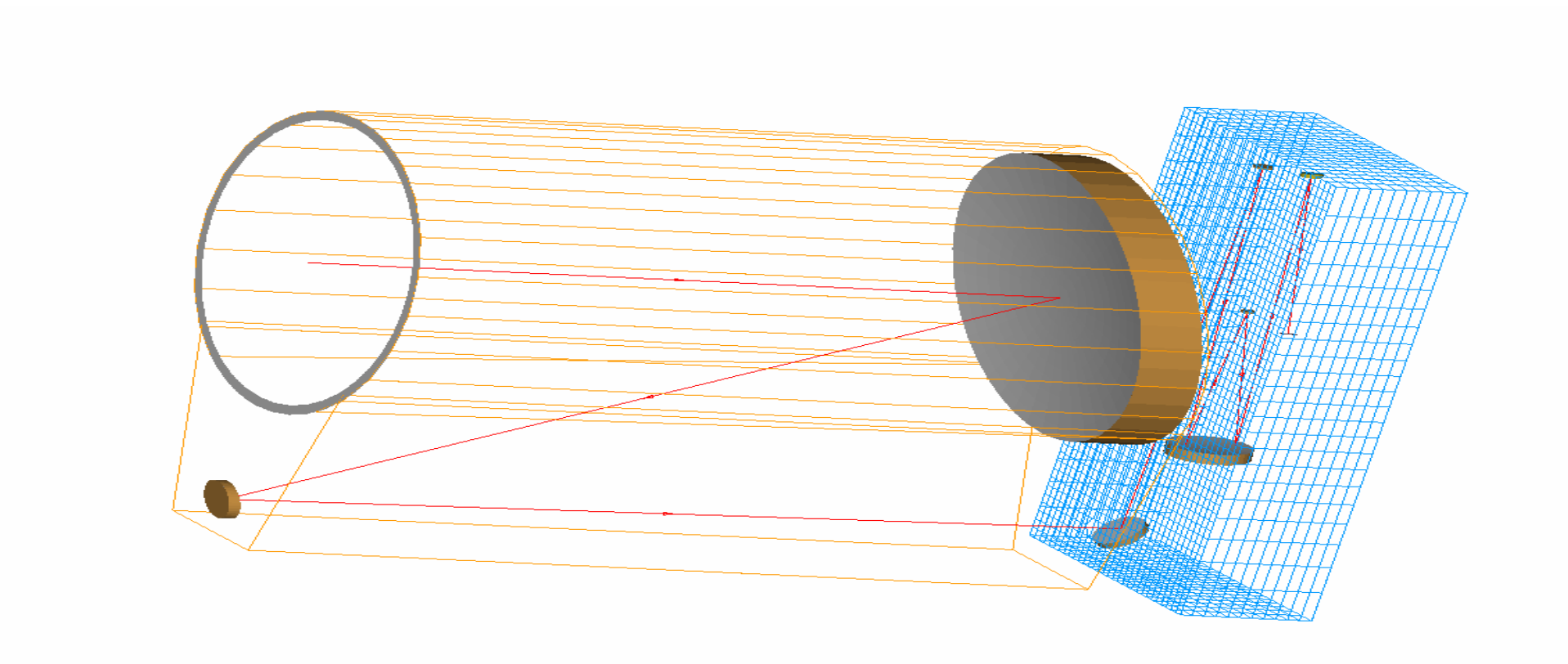

**Figure 2:** SEE-COAST off–axis parabolic primary mirror and FSM optics.

### 3.3 **Chromatic separation**

We choose to split the SEE-COAST operational wavelength range (0.4-1.25 micron) into 2 spectral channels covering the visible (0.4 – 0.85 micron) and the near-infrared (0.9 -1.25 ∝) essentially because coronagraphs have an efficient rejection factor only in a limited band-pass (it may be necessary to further subdivide the two channels for this reason). This corresponds to detector sensitivity ranges (CCD for the VIS and NICMOS for the near-IR).

### 3.4 **The coronagraph**

SEE-COAST instrumentalists from LESIA (Paris Observatory), IAGL (Liège) and JPL have developed a handful of different coronagraphic devices: classical Lyot amplitude coronagraphs with refined apodization schemes (Soummer 2005), refined Lyot coronagraphs (band-limited, Kuchner and Traub 2002), and various small inner working angle (IWA) phase-mask coronagraphs such as the four-quadrant phase-mask (FQPM, Rouan et al. 2000) and AGPM (Mawet et al. 2006). Brief descriptions and potential advantages are as follows.

**Four Quadrant Phase Mask (FQPM):** This divides the stellar PSF into four quadrants. In two of them (diagonally opposed) the wavefront is phase shifted by $\pi$, resulting in an interferometric nulling of the stellar peak. The FQPM has been intensively benchmarked through numerous lab experiments, in the visible (Riaud et al. 2003, Baudoz et al. 2007, in prep,), near-infrared (Mawet et al. 2006), mid-infrared for JWST/MIRI (Baudoz et al. 2005). Three FQPMs have already been installed at the VLT's adaptive optics system (Boccaletti et al. 2004, Riaud et al. 2006) and an intensive research & development on the problem of achromatization is ongoing. Several solutions have been proposed and tested in the lab (half-wave plate 4QPM and multi-4QPMs, see Figure 3).

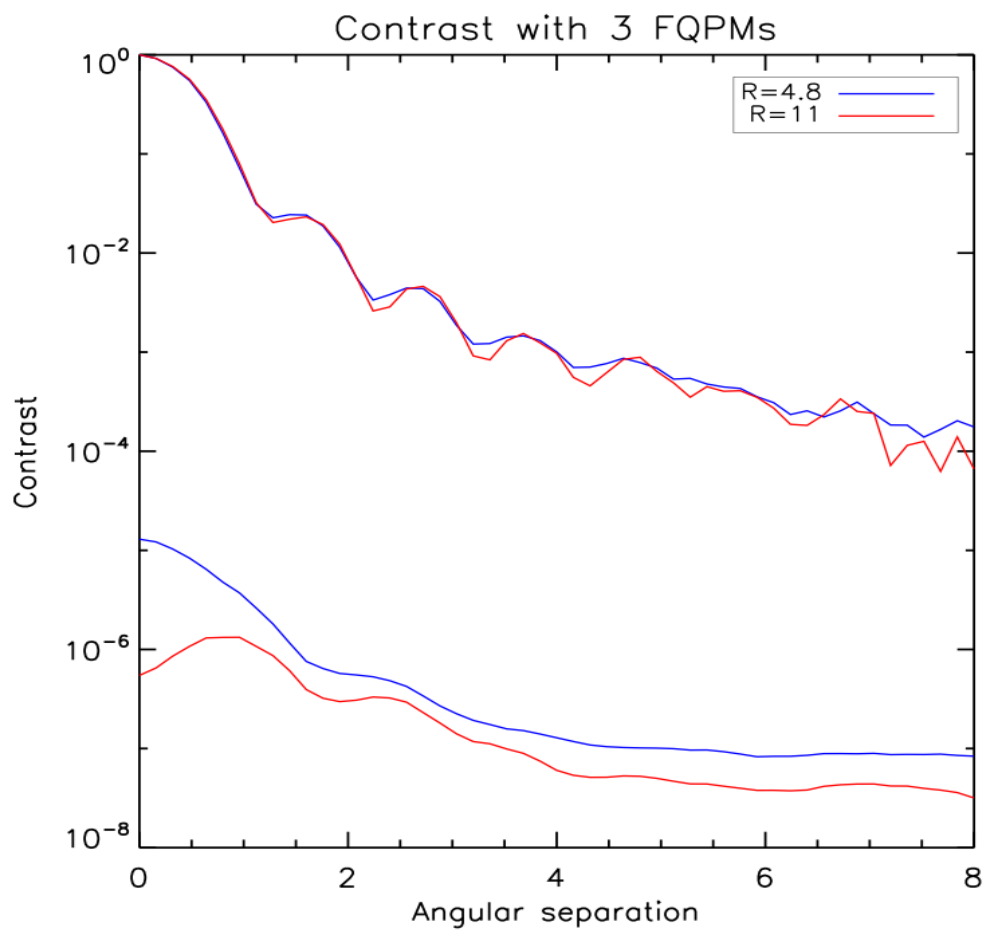

**Figure 3:** Lab result obtained with a cascade of 3 FQPMs in a 20%-bandwidth. A contrast of $10^{-6}$ has been measured as close as 2 λ/D without any active wavefront correction.

**Annular Groove Phase Mask (AGPM):** The AGPM, a natural evolution of the FQPM, consists of an achromatic optical vortex coronagraph of topological charge lp=2 (OVC2) manufactured thanks to integrated micro-optical elements (Mawet et al. 2005) (see Figure 4). Extensive analysis has assessed that a total null depth of about 5 x $10^{-6}$ (a contrast of ~ 5 x $10^{-7}$ at 2 λ/d) over 20% bandwidths can be obtained in natural unpolarized light, a result that can be improved by a factor 100 over 50% bandwidth if implemented in a spectro-polarimeter such that as considered for SEE-COAST. Apart from its very good small IWA and optical throughput, the AGPM also has a full discovery space (no dead zones). Higher order topological charges can be integrated (OVC4, OVC6), leading to improved tip-tilt/low-order aberration/stellar diameter sensitivity immunization.

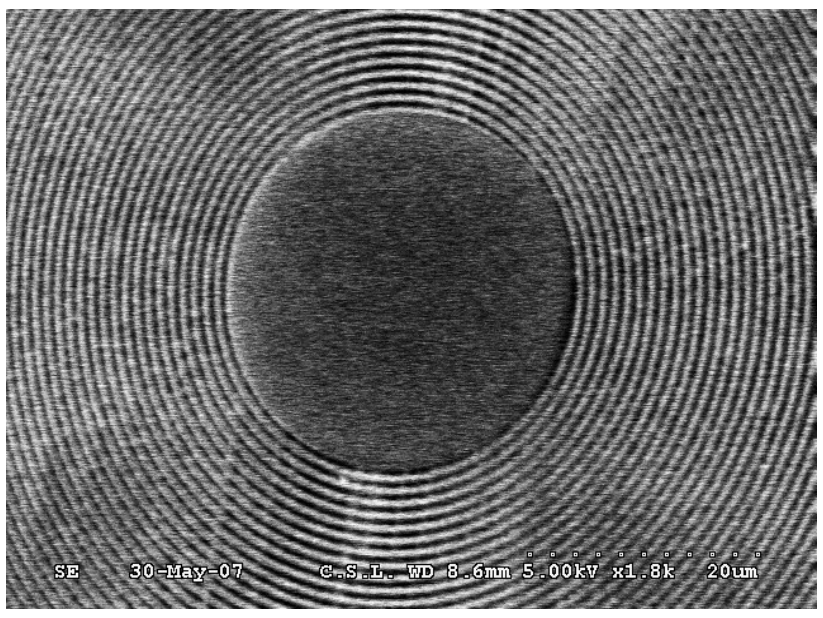

**Figure 4:**The AGPM consists of a concentric circular sub-wavelength grating, generating a vectorial optical vortex, i.e. a very efficient coronagraph. This SEM picture is an AGPM prototype manufactured at CSL.

**Band-limited Lyot (BL4):** This system uses a fuzzy bar-code pattern on a piece of glass to suppress the starlight diffraction pattern; in the theoretically perfect case, as with every other coronagraphs pre-selected for SEE-COAST, the remaining diffracted light background is zero (Kuchner and Traub, 2002). Note that, as every other amplitude coronagraphs, the BL4 throughput, IWA (> 3 $\lambda$/d) and discovery space is limited.

**Phase Induced Amplitude Apodizer (PIAA):** This uses a pair of warped mirrors to rearrange the intensity pattern of the incident starlight so that when this light is focused and the central star image blocked (thanks to a classical Lyot dot) the remaining diffracted light can, in the perfect case, be very low (Guyon, 2004). It has a good combination of transmission efficiency and angular resolution.

### 3.5 The low resolution spectrograph

SEE-COAST will be equipped with an Integral Field Spectrograph (IFS) in the visible and NIR. A key requirement for the IFS is to deliver a PSF data cube with the highest possible fidelity, i.e., one with speckle structure showing a high degree of correlation at all wavelengths. A TIGER-type IFS using a micro-lens array for image segmentation minimizes non-common aberrations. A scheme of such an IFS concept, directly inherited from the deveopments of the coronagraphic cameras SPHERE at the VLT (Beuzit et al 2006) and Gemini

Planet Imager (GPI) at the Gemini Telescope (McIntosh et al. 2006) , is shown in Figure 5.

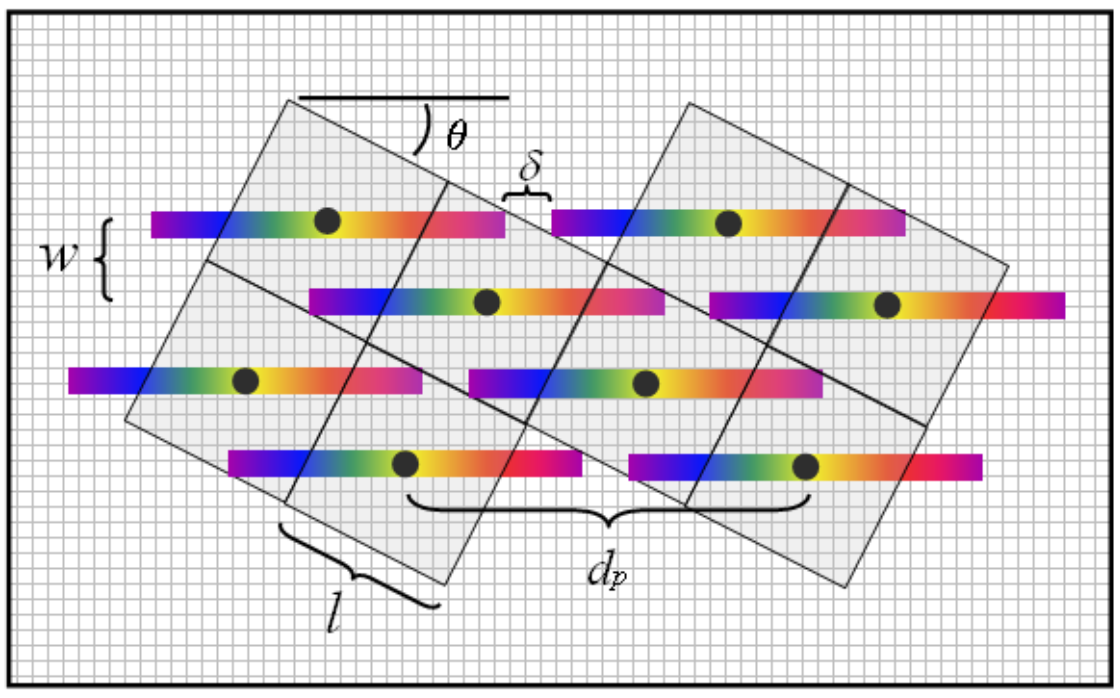

**Figure 5:** *S*chematic layout of a (Tiger-type) lenslet-based IFS.

Table 5: Requirements for an IFS in the visible.

| | |
|---|---|
| Detector | 2 x CCD cam 2Kx2K |
| Wavelength range | 0.4-0.65 / 0.65-0.85 micron |
| Spectral resolution | ~40 |
| Field of view | 3"x3" |
| Lenslet sampling | 18 / 30 mas/microlens[1] |
| Lenslet pitch | 50 micron (square) |

## 3.6 **The polarimeter**

SEE-COAST includes polarimetry as an integral part of the instrument, rather than as an add-on. The design and specification of the polarimetry optics is thus in the baseline. It is also important that the design of other optics, particularly those upstream of the polarization modulator, minimizes any instrumental polarization, which is crucial for the required polarization sensitivities (~$10^{-5}$ , which is the current state-of-the-art with SPHERE, the “Extreme Polarimeter” camera (ExPo - Keller 2006), and the PlanetPol polarimetric camera (Hough et al 2006). The baseline only includes polarimetry for the optical channel (0.4-0.85 µ), where linear polarization produced by scattering in the planetary atmosphere will be largest.

In addition, the polarization properties of the AGPM coronagraph make it especially suited to differential polarimetric imaging provided it is properly implemented inside the polarimeter. Its symmetry also allows for alternate switching between two orthogonal polarization states that are linear at the input.

See Fig. 6 for a shematic layout. In conclusion, a polarimetric system which converts the polarization signal using a polarization modulator into a temporal intensity modulation can be designed to accommodate the AGPM coronagraph, the mutual advantages of which lead to a substantial performance improvements. In space, the polarimetric modulation can be much slower than for ground-based polarimeters such as the Zürich IMaging POLarimeter (ZIMPOL – Gisler et al. 2004) schematically shows the coronagraphic spectro-polarimeter. More classical polarimeter implementations will also be considered in collaboration with SPHERE, PlanetPol, and ExPol polarimetrists.

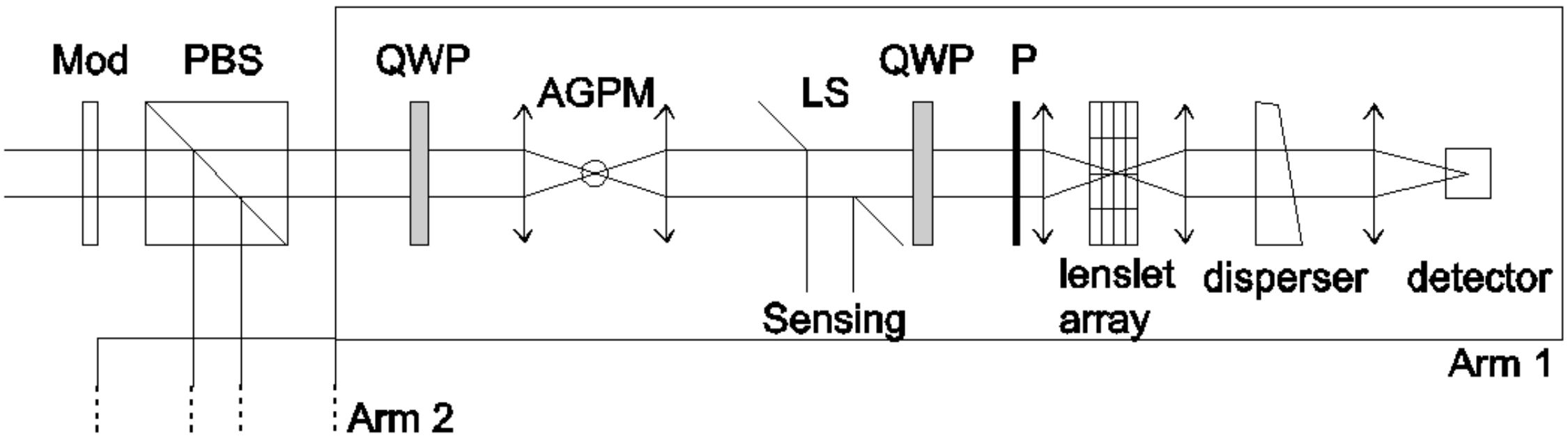

**Figure 6:** Optical schematic of an implementation of SEE-COAST's coronagraphic imaging spectro-polarimeter. The beam is first slowly modulated by a variable retarder (Mod, e.g. a rotating half-wave plate), then split by a polarizing beam-splitter (PBS) into 2 arms of orthogonal linear polarizations (Arm 1 and 2). The QWPs are achromatic quarterwave plates that convert linear to circular polarization to efficiently feed the AGPM coronagraph. LS stands for Lyot stop. The IFS is based on a TIGER-type lenslet-array.

### 3.7 **Wavefront sensing and control**

In the case of the second option described in section 3.2 (introduction of a deformable mirror), wavefront sensing and correction is needed to provide high contrast images. We propose to carry out this wavefront analysis in two steps: first a “classical” analysis to reduce the impact of low-order aberrations and then a careful analysis of the detector image to cancel out the spurious speckles. The correction is achieved with at least one Deformable Mirror (DM) and ideally two.

**1st stage** The purpose of this first analysis is to measure the first optical modes with an accuracy of a few nanometres and a repeatability less than or about one nanometre rms. In particular, tip-tilt has a dramatic impact on the coronagraphic performance, and the low-order aberrations must be known to actively control the telescope alignment. The first analysis can be made with standard methods involving either pupil plane measurements (a well mastered technique but susceptible to non-common path aberrations) or focal plane measurements (phase diversity principle). The latter appears more suited to the context of SEE-COAST since aberrations are expected to evolve slowly. Both solutions have already been implemented in several space or ground based systems and have demonstrated their performance and reliability. This 1st stage might not be needed if the 2nd alone stage can deal alone with the amount of phase aberrations in SEE-COAST, a likely situation given the environment and the specifications made on the optical system.

**2nd stage** Standard wavefront control is not sufficient for SEE-COAST and a second stage is thus desirable. The purpose of this stage is to measure directly the residual phase aberrations in the coronagraphic image to take advantage of all the available information. The intensity distribution in a coronagraphic image is related to the square modulus of the Fourier Transform of the phase. An appropriate algorithm can be used to recover the phase and apply the desired level of correction. The expected gain is very large, a factor of 100 in wavefront rms, and a factor of 10,000 in reduction of focal-plane speckles. This translates to a potential situation in which an exoplanet goes from being buried under 1000-times brighter speckles to one in which the exoplanet stands out above the background speckles by a factor of 10. This is illustrated in a paper by Trauger and Traub (2007), which used a laboratory version of a space coronagraph system, very similar to the focal-plane sensing concept of SEE-COAST. If one DM is used, half of the field of view of a coronagraph can have its speckles reduced; both phase and amplitude non-uniformities in the wavefront are minimized in this half ‘dark hole.’ If 2 DMs are used, sufficiently separated to use the Talbot effect, the full field of view can be made into a dark hole. This is the rationale for desiring 2 DMs per output channel.

### 3.8 **Speckle calibration**

Alternatively, or in combination to wavefront control speckle calibration is envisaged. The performance of the coronagraph is limited by its intrinsic defects and the wavefront errors introduced by the telescope and the relay optics. Most of the polishing and coating defects will be corrected by the DMs. The residual uncorrected errors scatter light in a speckle pattern. The expected level of the speckles at 2-4 $\lambda$/D is about $10^{-7}$ to $10^{-8}$. To reach the goal of detecting a planet with a contrast of $10^{-9}$ at 2 $\lambda$/D, we will need to extract the planet image from residual speckles 10 to 100 times brighter than the planet itself. To do so, a dedicated calibration strategy has to be implemented inside SEE-COAST. Calibrating the speckles simultaneous to the scientific detection appears to be the safest option. Besides the well-proven *polarization differential imaging (PDI) and spectral differential imaging (SDI)* techniques that are inherent to the SEE-COAST instrumental concept, we describe below non-exclusive options that could be used in to improve the calibration by the spectro-polarimeter.

**Option 1: Self-Coherent Camera.** One solution is to calibrate the speckles using a technique based on the lack of coherence between the stellar and planetary emissions (Galicher & Baudoz. 2007). The Self-Coherent Camera splits the beam in two, one used as a coherent reference for the other. The beams are recombined in a Fizeau scheme. Thus in the recorded image, the speckles are encoded with fringes while the planet beam is almost unaltered. A data reduction algorithm is able to discriminate fringed speckles from the un-affected planet.

**Option 2: Pupil Remapping** This option is to use a remapping system, as described in Perrin et al. (2006). The pupil located after the coronagraph is broken into subpupils each injected into single-mode fibres to perfectly spatially filter the beams. These subpupils are re-arranged into a non-redundant output pupil to avoid mixing phases from different parts of the input pupil. Since it allows self-calibration of the Optical Transfer Function, the remapping system allows us to distinguish instrument perturbation from astrophysical information.

## 3.9 The detectors

The detectors specified for SEE-COAST are classical, low-readout noise devices.

**Optical CCD detectors** The baseline optical detector is the CCD47 series of Charge Coupled Devices (CCD), which have 1k x 1k pixels each of 13.5 µm square and are manufactured by E2V Technologies, U.K. This family of devices has a long and successful heritage in space missions and come in custom, high performance space packages. They are radiation tolerant devices which can be optimised with specific structures and processes that are backed up by detailed theoretical models and experimental data on radiation damage effects. The CCDs are sensitive from 250 nm to 1000 nm, have typical readout noise of less than 3 e- rms, and dark current rates of less than 0.003 e/pixel/s at operating temperatures of 170K.

**Near Infrared Detector** The baseline Near Infrared detector is Teledyne Scientific (formerly Rockwell) short wave cut-off HgCdTe (grown on CdZnTe but with the substrate removed) 1k x 1k array bump-bonded onto a Hawaii 1RG multiplexer (18um square pixels). This detector is sensitive from 0.5 to 1.7 µm and will have better than 25 e- rms read noise and 0.02 e-/s/pixel dark current at an operating temperature of 120-140K. The HAWAII*RG detector family is a fully developed product with a stable manufacturing process and therefore requires no new development, having been developed specifically for the JWST. In the standard 100 kHz readout mode, the full array can be read out in less than 680ms using 16 outputs or 5s using 2 outputs.

### 3.10 Calibration and data processing

**- Detectors:** A very accurate Flat Field determination is mandatory especially when dealing with differential imaging, where the various images are recorded at different locations on the detector. From previous studies (SPHERE) we anticipate that an accuracy of about $10^{-4}$ would be required. In any case, the integration time to achieve such a level on a Flat Field calibration amounts to something like 1500s and would therefore be a significant part of the calibration procedure. The situation is different for a system based on the ZIMPOL concept, in which the differential images are actually recorded on the same pixels therefore relaxing the Flat Field constraints.

**- Spectrograph:** The calibration will be done with spectra of “standard” stars.

**- Instrumental polarization:** Polarization calibration is needed because of the oblique reflections in front of the polarimeter as well as the non-perfect polarimeter optics themselves. Fortunately, the required polarization accuracy (accuracy with which a very small polarization is measured) is only 10%, as compared to the required polarization sensitivity (smallest amount of polarization that can be detected) of $10^{-5}$. For comparison, the solar spectro-polarimeter on the Japanese Hinode satellite is calibrated to better than 1%. Therefore, fairly simple calibration approaches are sufficient to meet the requirements.

# 4.Expected performance

Numerical simulations have been started to address the performance of SEE-COAST on the basis of simple hypothesis. These results are presented here. A sensitivity analysis involving a more complete numerical model is now required to define the requirements for achieving the final level of contrast.

We have simulated the planet detection performances with a specific designed software called *CoastSim* developed within our team. It uses the following realistic hypothesis:

- Sstandard stellar and planetary spectra for giant planets and super-Earth
- Finite stellar radius
- System characteristics:
  - WFE: lambda/100 rms @ 633 nm (Power Spectrum Density: $f^{-1.5}$)
  - Throughput: 25%
  - Jitter: 0.5 mas rms / 3 axis
  - Quantum efficiency = 80% CCD / 55% HgCdTe
  - Readout noise : 2e- CCD / 15 e- HgCdTe
  - Dark current: 0.001 e-/s
- Coronagraph characteristics:
  - Realistic mask imperfections in phase and amplitude.
  - WFE after the coronagraph (lambda/100 rms)
  - Non-common path error: 0.2 nm rms
- Background noise:
  - Solar System zodiacal light
  - Exo-zodiacal light (10 x zodi)
- Speckle noise (after calibration): the simulations were done with the AGPM coronagraph (Figure 7).

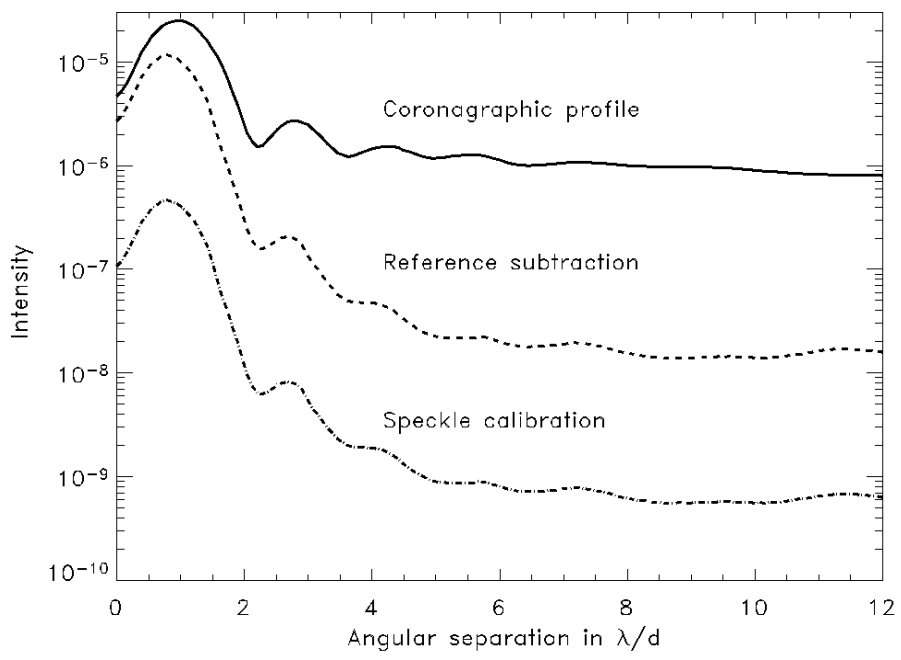

Figure 7: AGPM coronagraphic profiles (raw, after reference subtraction, after speckle calibration) according to CoastSim and the hypothesis presented in the text.

Figure 8 and Figure 9 show the number of detections by SEE-COAST assuming that each of the 129 targets chosen for this simulation (including potential target stars for SEE-COAST) possesses one Earth-like planet (broad band imaging, bandwidth 20%). Figure 10 is identical for a spectroscopic resolution of R=40.

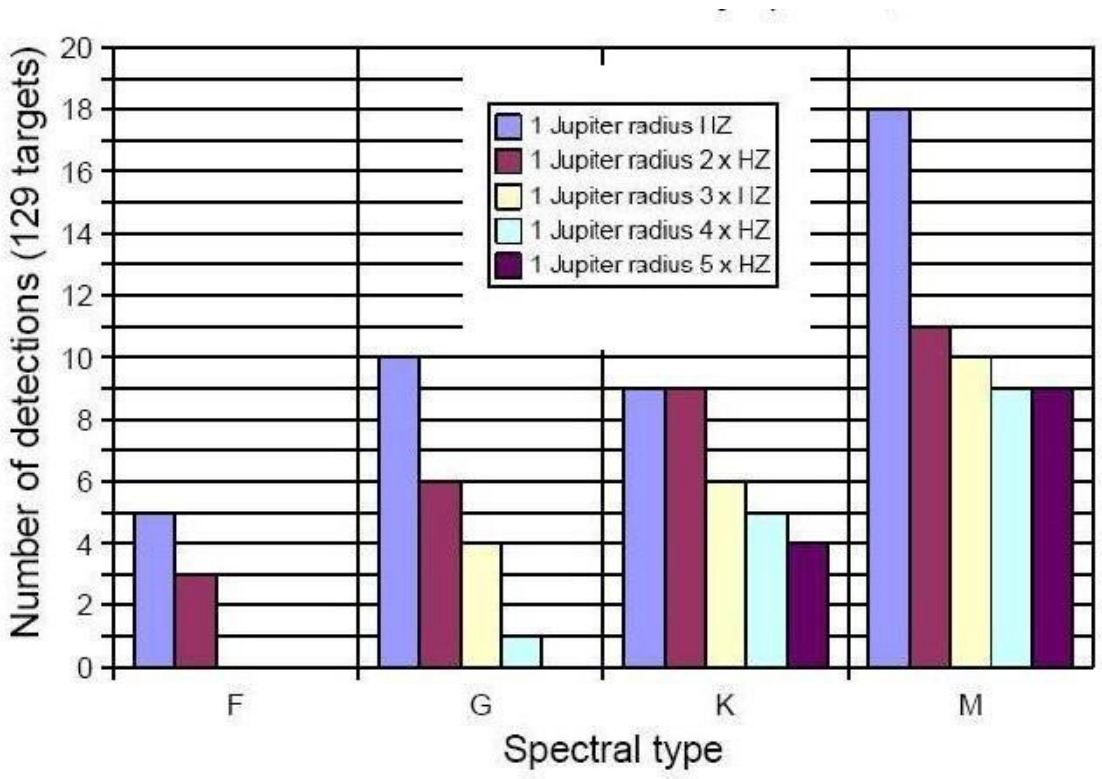

Figure 8: Statistics of Jupiter-detection (bandwidth 20%) and SNR>5 (1 day).

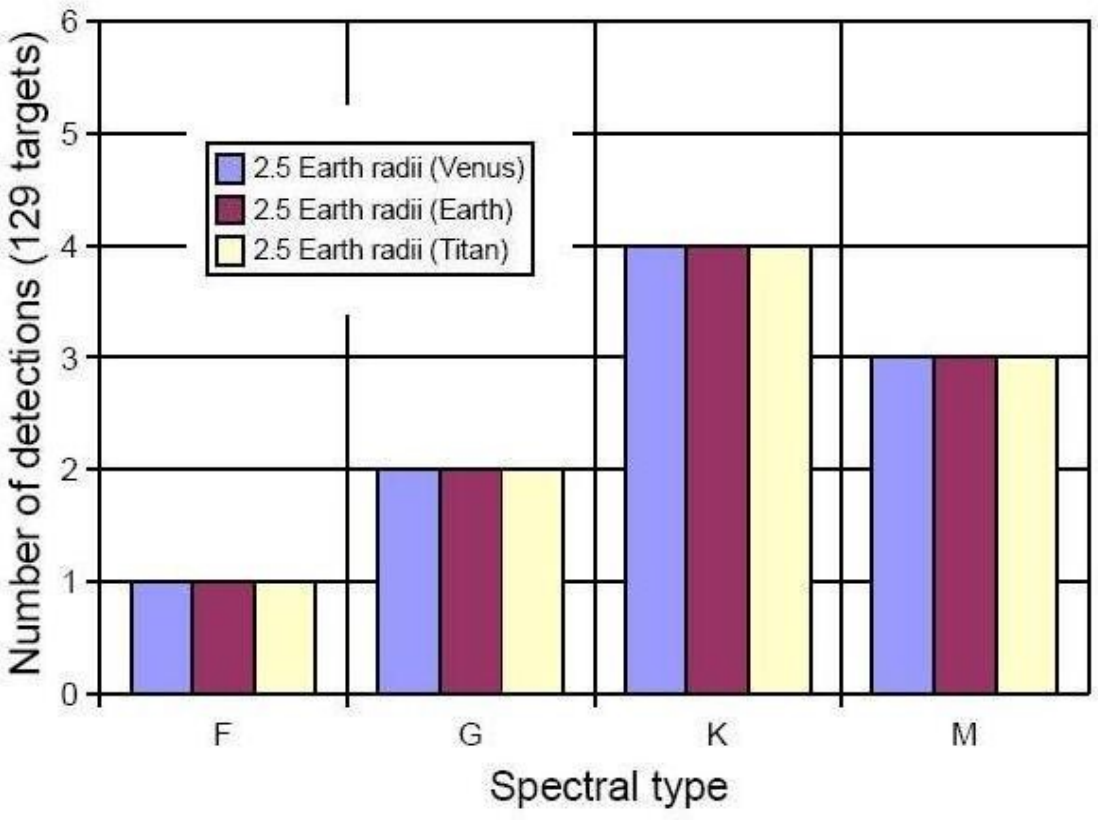

Figure 9: Statistics of Super-Earth detection (bandwidth 20%) & SNR>5 (3 days exposure).

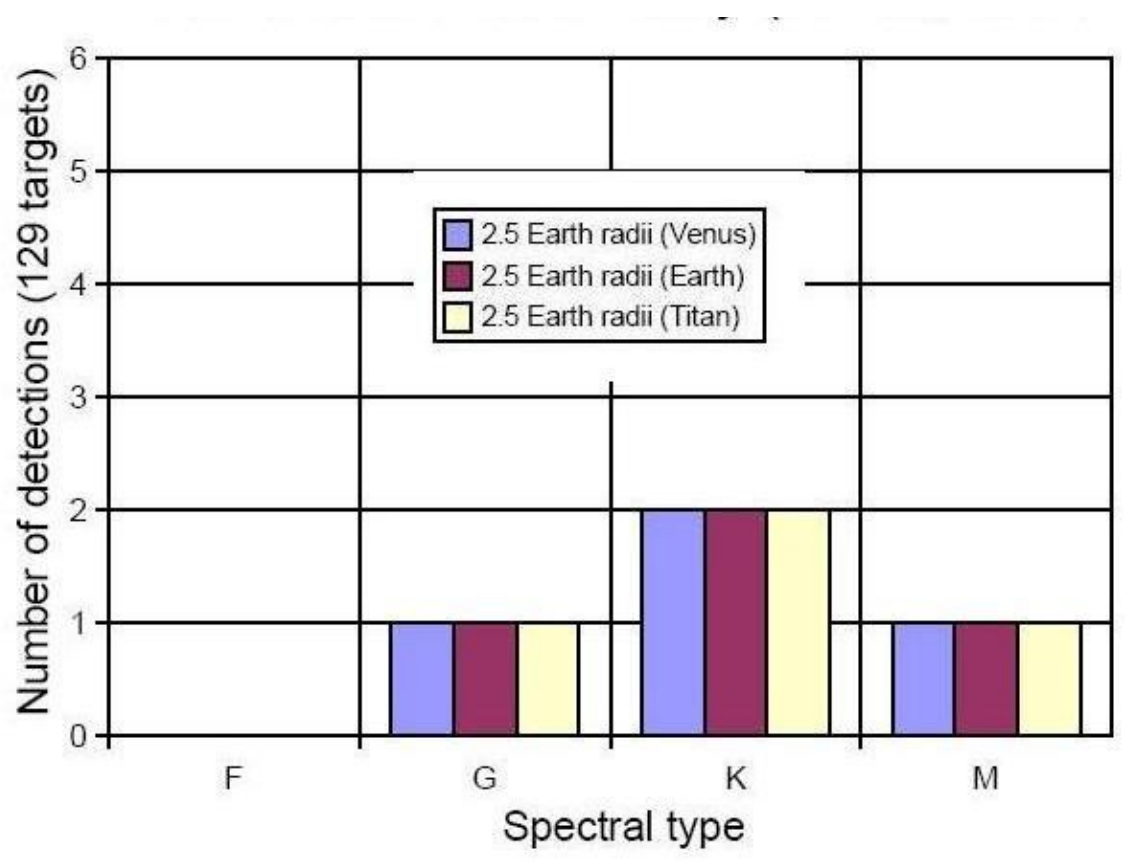

Figure 10: Statistics of Super-Earth detection @ R=40 & SNR>5 (3 days exposure).

Figure 11 and Figure 12 show the R=40 spectra obtained after spectral calibration for Super-Earth and Super-Titan objects assuming 3 days of exposure time.

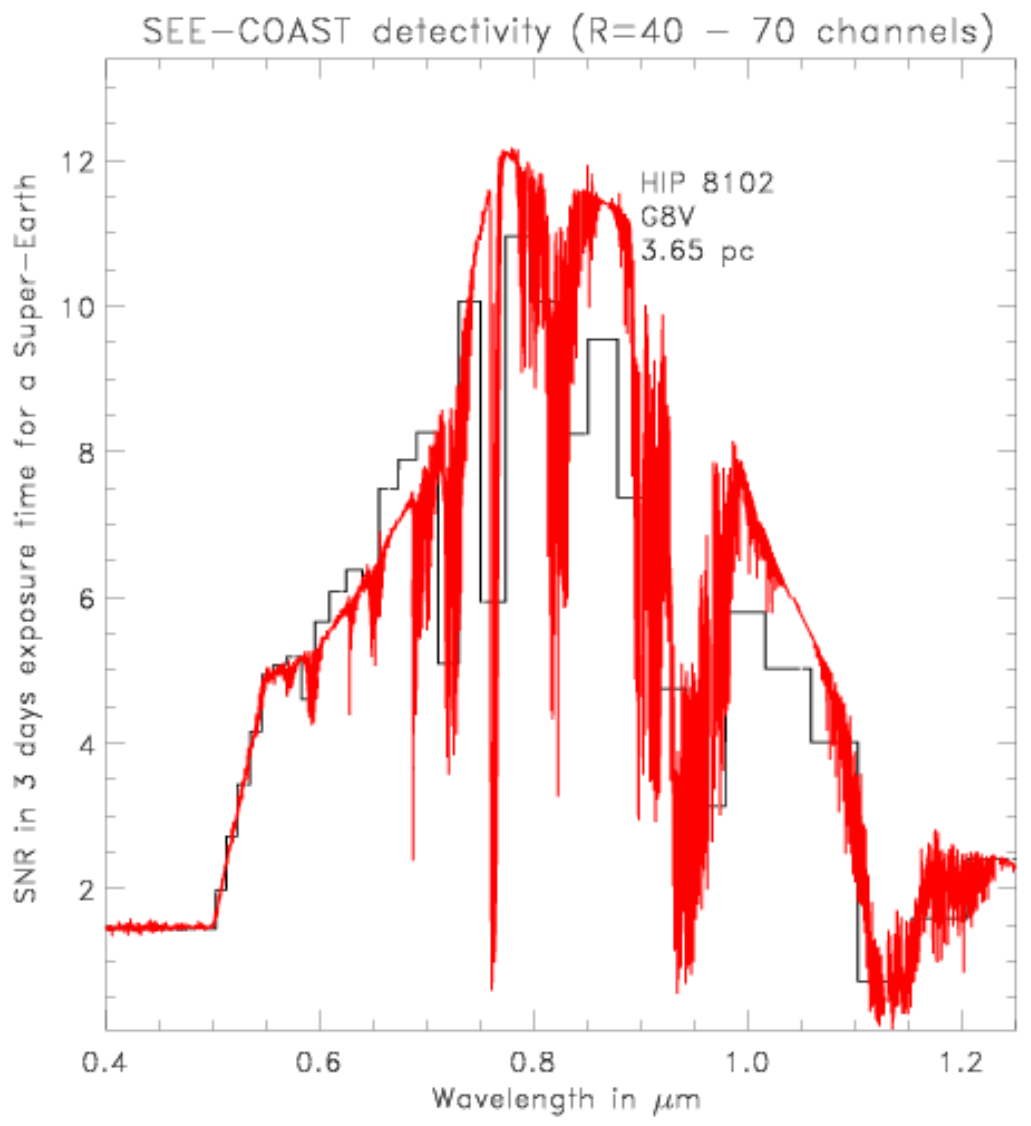

Figure 11: A spectrum of a Super-Earth planet (2.5 Earth radii) in the habitable zone, seen with R~40 spectral characterization by SEE-COAST (3 days exposure).

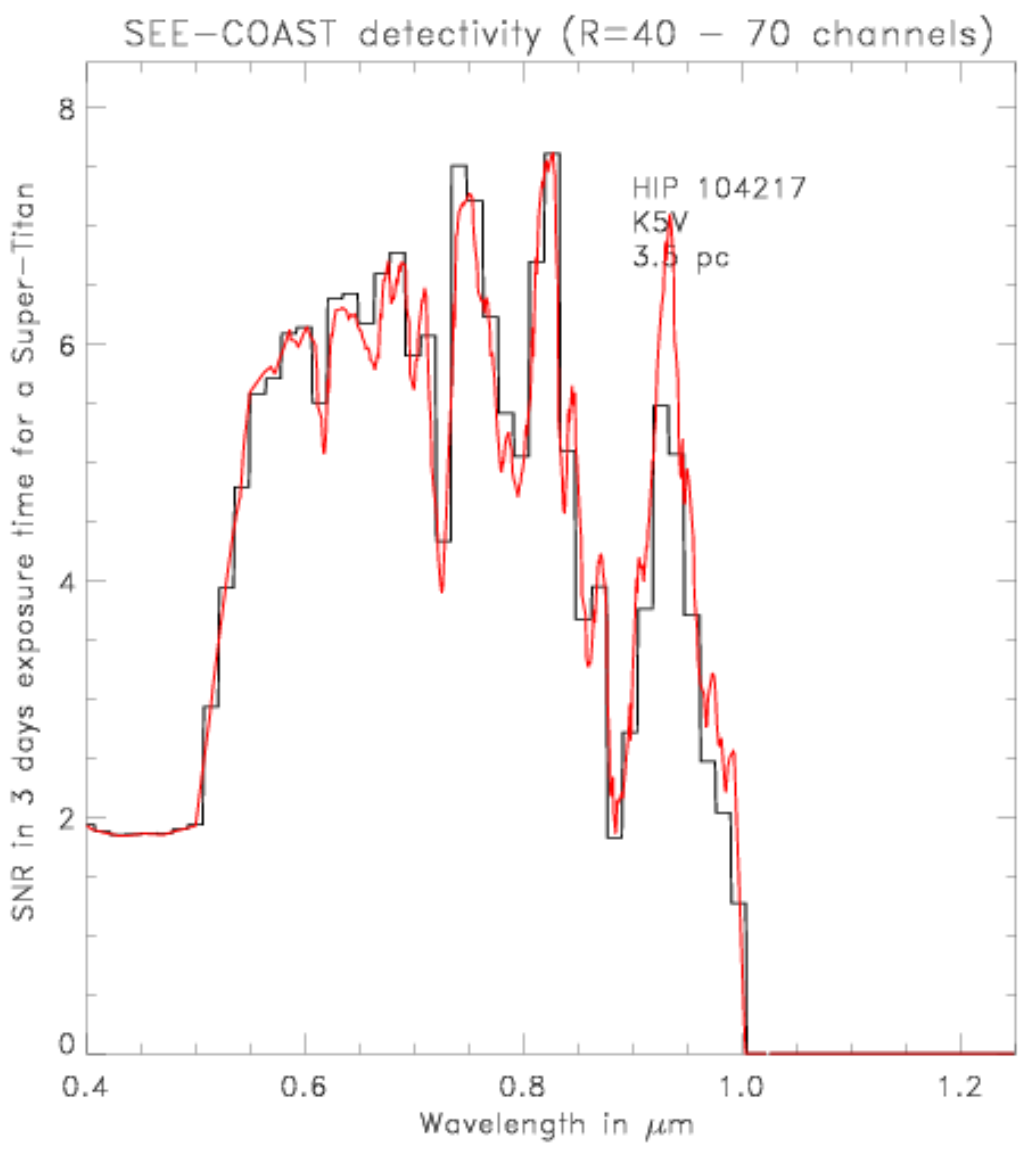

Figure 12:A Super-Titan planet in the habitable zone spectral characterization by SEE-COAST (3 days exposure).

In the example shown in Figure 11, the SNR per channel is typically 8 in a 3 days exposure. With this performance, a degree of polarization (relative difference for 2 polarization states) of 40% can be measured for a super Earth with SNR=4 and for a Jupiter with SNR=12.

# 5.Conclusion

A coronagraphic mission like SEE-COAST is an interesting precursor mission to carry out the physico-chemical characterization of giant planets and some super-Earths before more ambitious missions like large coronagraphic (4-8m class) missions, large interferometers, or external occulters. In addition, its reflected light approach makes it complementary to missions aimed at the thermal emission of exoplanets (see Schneider 2003 for a visible-infrared comparison).

# 6.References

Baudoz P., Boccaletti A., Riaud P., Cavarroc C., Baudrand J., Reess J.-M. & Rouan D., 2007., PASP , 118 , 765

Beauliau J.-Ph., Bennett D., Fouqué P. et al., 2006., Nature, 439, 437

Benz W., 2006. Meteoritics & Planetary Science, 41, #5393

Beuzit J.-L., Feldt, M., Mouillet D., Moutou C., Dohlen K., Puget P., Fusco T., Baudoz P., Boccaletti A., Udry S., Ségransan D., Gratton R., Turatto M., Schmid H.-M., Waters R., Stam D., Rabou,P., Lagrange A.-M.; Ménard F., Augereau J.-C., Langlois M., Vakili F., Arnold L., Henning T., Rouan D., Kasper M. and Hubin N. 2006. in IAU Colloquium 200. p. 317

Cockell Ch., Herbst T. and 40 co-authors. 2008. Experimental Astronomy in press.

Fortney J., Marley M. & Barnes J., 2007. ApJ. 659, 1661

Fortney J., Saumon D., Marley M. et al., 2006. ApJ., 642, 495

Galicher R. & Baudoz P., 2007. C.R. Acada. Sci. Paris (Physique) , 8 , 333

Gisler D., Schmid H.-M., Thalmann Ch., Povel H.-P., Stenflo J., Joos F., Feldt M., Lenzen R., Tinbergen J., Gratton R., Stuik R., Stam D., Brandner W., Hippler S., Turatto M., Neuhauser R., Dominik C., Hatzes A., Henning Th., Lima J., Quirrenbach A., Waters L., Wuchterl G. and Zinnecker H. 2004. in SPIE, Volume 5492, p.. 463- Eds. Alan F. M. Moorwood and Iye Masanori

Guyon O., 2004. ApJ , 615 , 562

Ida S. & Lin D., 2004., ApJ, 604, 388

Hough J., Lucas P., Bailey J., Tamura M., Hirst E., Harrison D. and Bartholomew-Biggs M. 2006. in Publications of the Astronomical Society of the Pacific, Vol. 118, p.1302

Keller C.U., 2006, Design of a polarimeter for extrasolar planetary systems characterization, SPIE 6269, 62690T

Kemp J., Henson G., Steiner C. & Powell E., 1987. Nature, 326, 270

Kuchner M., 2003. ApJ. Letters, 596, L105

Kuchner M. & Traub W.? 2002. ApJ. , 570 , 900

Léger A., Selsis F., Sotin Ch. Et al., 2004., Icarus, 196, 499

Mawet D., Riaud P., Absil O. & Surdej J., 2006. ApJ. , 633 , 1191

Mayor M., Bouchy F., Benz W., Lovis Ch. ,Mordasini Ch., Pepe F., Queloz D., Udry S. et al. 2008. In “Extrasolar Super-Earths”. Colloqium held in Nantes (June 2008).

McIntosh B., Graham J., Palmer D., Doyon R., Larkin J., Oppenheimer B., Saddlemeyer L., Veran J., Wallace J. &Gemini Planet Imager Team , 2007. BAAS, 39 , 030.05

Mordasini ch., Alibert Y., Benz W. and Naef D., 2007. in Extreme Solar Systems, ASP Conference Series, eds. Fischer D., Rasio F., Thorsett S. and Wolszczan A.

Palle E., Ford E., Seager S., Montanez-Rodriguez P. & Vasquez M. 2008. ApJ. , 676 , 1319

Perrin G., Lacour S., Woillez J. & Thiebault E., 2006. MNRAS , 373 , 747

Riaud P., Boccaletti A., Baudrand J. & Rouan D., 2003. PASP , 115 , 712

Rivera E., Lissauer J., Butler P. et al., 2005., ApJ,, 634, 625

Rouan D., Riaud P., Boccaletti A., Clenet Y. & Labeyrie A., 2000. PASP , 112 , 1479

Sato B., Fischer D., Henry G. et al. 2005., ApJ., 633, 465

Schneider J., 2003. In Toward other Earths. Darwin/TPF and the Search for Extrasolar Terrestrial Planets. ESA SP-539.

Schneider J., 2008. Extrasolar Planet Catalog. http://exoplanet.eu

Seager S., Whitney B. & Sasselov D., 2000. ApJ., 540, 504

Sotin Ch., Grasset O. & Mocquet A., 2007., Icarus, 191, 26

Soummer R., 2005. ApJ. Letters , 618 , L161

Sozzetti A., Casertano S., Lattanzi M., Spagna A., Morbidelli R., Pannunzio R., Pourbaix D. and Queloz D., 2007. in IAU Symposium 248 - A Giant Step: from Milli- to Micro-arcsecond Astrometry

Stam D., Hovenier J. & Waters L., 2004. Astron. & Astrophys., 428, 663

Swain M. Vasisht G. & Tinetti G., 2008, Nature, April

Tinetti G., Vidal-Madjar A., Liang M.-C. et al., 2007., Nature, 448, 169

Trauger J. and Traub W., 2007. Nature, 446, 771

Udry S., Bonfils X., Delfosse X. et al., 2007., Astron. & Astrophys., 469, 43

Valencia D., Sasselov D. & O'Connel R. 2007., ApJ. 665, 1413

Williams E. & Gaidos E. 2008. Icarus, 195, 927

Wolszczan A. & Frail D., 1992, Nature, 335, 145

The SEE-COAST Team (by alphabetic order of country):

J. Schneider (PI), R. Dvorak, S. Lacour, Ch. Hanot, D. Mawet, S. Roose, Y. Stockman, J. Surdej, M. Beaulieu, R. Doyon, R. Jayawardhana, D. Lafrenière, J.-F. Lavigne, C. Marois, J. Matthews, B.Netterfield, N. Rowlands, M. van Kerkwijk, J.-P. Veran, B. Merin, M. Barthelemy, P. Baudoz J.-L. Beuzit, A. Boccaletti, T. Fusco, O. Grasset, J.-M. Grießmeier, A. Laurens, Th. Lépine , B. Lopez, F. Ménard, A. Morbidelli , L.Mugnier, G. Perrin, D. Rouan, H. Michaelis, H. Rauer, S. Wolf, A. Sozzetti, O. Guyon,P. Montanes-Rodriguez, E. Palle,Y. Alibert, W. Benz, H.-M. Schmid, S. Udry, P. Ehrenfreund, J. Hovenier, C. Keller, A.Selig, F. Snik, D. Stam, A. Aylward, J. Cho, R. Cole,, P. Doel, J. Hough, D. Ives, A. Longmore, G. Tinetti, R. Angel, R. Brown, G. Laughlin, G. Marcy, M. Marley, I. De Pater, L. Kaltenegger, J. Kasdin, J. Kasting, D. Sasselov, S. Seager, W. Traub, J. Trauger & Y. Yung